# Addressing limitations of the endpoint slippage analysis

Marco-Tulio F. Rodrigues

Chemical Sciences and Engineering Division, Argonne National Laboratory, Lemont, IL, USA

**Contact:** marco@anl.gov

**Abstract**

Some rate of oxidation and reduction side-reactions will inevitably coexist in most rechargeable batteries, contributing to reversible and irreversible self-discharge. While parasitic reduction traps electrons, parasitic oxidation donates electrons to the cell's inventory and can lead to temporary capacity gain. This causes direct capacity measurements to be an unreliable source of information about the total extent of side-reactions happening in the cell. The most widely used method to determine the rate of these two types of parasitic processes involves analyzing the slippage of endpoints, which consists in tracking the termination of cell charge and discharge when data is represented along a cumulative capacity axis. Here, we argue that this approach could lead to inaccuracies when applied to certain systems, which includes Si electrodes in Li-ion batteries and hard carbon in Na-ion batteries. We analyze this issue in quantitative terms and propose equations that can provide true rates of parasitic processes from experimental endpoint slippage data. This work shows that, in battery science, well-established analytical approaches may not be directly transferrable to new electrode systems.

**Introduction**

Li-ion batteries always experience some rate of side-reactions. These parasitic processes will typically consume electrolyte species, which ultimately leads to performance degradation. Typical examples of such processes are the electrolyte reduction at the negative electrode (NE) and oxidation at the positive electrode (PE).[1-3] The former leads to formation and maintenance of the solid electrolyte interphase (SEI), while the latter causes reversible self-discharge during extended storage at high voltages.

The directionality of the electron transfer during these parasitic reactions determines their possible consequences to cell capacity. During reduction reactions associated with the SEI, electrons are abstracted from the NE at an average rate $I_{red}$ and irreversibly consumed in the formation of insoluble components. This depletes the initial electron inventory contained within the PE and can decrease cell capacity. During electrolyte oxidation, electrons are donated by the reacting species (e.g., ethylene carbonate) at an average rate $I_{ox}$ to the PE, where they combine with a $Li^+$ from the electrolyte. This adds to the initial electron inventory brought into the cell by the PE and can lead to an increase in cell capacity.[1-3] Both these examples involve the destruction of solvent species, which can eventually lead to de-wetting of electrode domains and an increase in cell impedance.[4] These secondary consequences, however, may only manifest after extensive aging and, until that happens, parasitic oxidation can offset some of the capacity losses that would otherwise be observed due to SEI growth. Consequently, capacity measurements per se are not always reliable sources of information about the rate of aging experienced by the cell. This becomes particularly problematic when screening new electrolyte compositions, as parasitic oxidation could run undetected in short-term tests.

Quantifying the capacity exchanged in parasitic reduction and oxidation is a task that can elude even detailed methods of data analysis. Differential voltage analysis and related approaches are among the most successful in extracting aging information from experimental data. These techniques consist in combining the voltage profiles of unaged electrodes to generate simulated full-cells.[5-8] The voltage profiles of electrodes can be modified (e.g., "squeezed" to simulate active material loss) until the dV/dQ or dQ/dV of the simulated cells matches the curves observed experimentally. This useful technique can quantify what is usually called "loss of Li$^+$ inventory" (LLI), which is inferred from changes in the offset between the voltage profiles of the PE and the NE needed to emulate the experimental data.[5, 6] LLI essentially represents the *net change* in electrons in the cell due to parasitic processes. Hence, these methods can only obtain $I_{red} - I_{ox}$, when perhaps $I_{red} + I_{ox}$ would be a better metric to gauge changes to the long-term health of the cell.

One way to quantify individual rates of parasitic oxidation and reduction has been proposed by Tornheim & O'Hanlon.[3] Their formalism shows that, under certain conditions, the deviation of coulombic efficiency from unity depends only on $I_{red}$, while capacity retention depends on both $I_{red}$ and $I_{ox}$. From these quantities (capacity and efficiency) it thus becomes possible to estimate the rates of both side-reactions at every cycle. Although this approach is simple and only requires quantities readily available from experiments, we are unaware of its use in the literature for this purpose. We have previously generalized this approach to be compatible with other conditions, as detailed in refs. [1] and [2].

The most used method to independently quantify $I_{red}$ and $I_{ox}$ involves analyzing the "slippage" of charge and discharge endpoints in full-cell voltage profiles plotted along a cumulative capacity scale (Figure 1).[a][9, 10] Charging the cell adds to the cumulative capacity, while discharging will subtract from it. For each cycle, the endpoints for charge and discharge are the values in the cumulative capacity axis at which the relevant half-cycle ends. This clever change in representation allows for separation of $I_{red}$ and $I_{ox}$. As we discuss below, slippage of the discharge endpoint (that is, its movement towards higher cumulative capacity values) is uniquely correlated to parasitic reduction, while slippage of the charge endpoint occurs as a consequence of parasitic oxidation. This method has been applied to several systems to provide a qualitative assessment of electrolyte reactivity under diverse experimental conditions.[9-21]

Consider Figure 2a, which shows voltage profiles for a $LiNi_{0.8}Mn_{0.1}Co_{0.1}O_2$ (NMC811) PE and a graphite (Gr) NE, plotted along the initial state of charge (SOC) scale of the PE. The initial lateral offset between the profiles (solid lines) represents electron losses during formation cycles. Since cell voltage is the difference between the electrical potentials of the PE and NE, these profiles define a full-cell. Parasitic reduction reactions will deplete electrons from the negative electrode without directly affecting the $Li^+$ content of the PE. That is equivalent to moving the NE profile away (dashed blue line) from the PE curve, so that a given delithiation state of the PE will now be associated with a lower $Li^+$ content at the NE. The same can be done during cell discharge (Figure 2b) to represent side-reactions occurring during that half-cycle. Finally, we can obtain the full-cell profiles by subtracting the NE from the PE at the two simulated cycles and applying the voltage limits, resulting in Figure 2c. Note that the curves in Figure 2c have been adjusted to be represented along a cumulative cell capacity axis, and clearly show discharge endpoint slippage as a response to parasitic reduction.

---

[a] Note that, in previous works (refs. [1] and [2]), we used the term "slippage" to refer to changes to the offset between the voltage profiles of the PE and NE (as discussed in Figure 2), which is different than the concept of slippage discussed here.

A similar representation can be done for oxidation. Here, oxidation will donate electrons to the PE without directly changing the electron content of the NE, which can be represented by sliding the positive electrode profile towards that of the NE (Figure 2d,e). Inspecting the simulated cell profiles (Figure 2f), it can be seen that only charge endpoint slippage occur as a result of parasitic oxidation.

The examples above demonstrate how this technique can distinguish $I_{red}$ from $I_{ox}$. This distinction is possible due to how the termination to each half-cycle is met. At the end of charge (EOC), Gr presents a plateau while the PE curve marches towards high potentials. In this case, the cell will reach its charge voltage cutoff depending on how quickly the PE potential increases, and hence, charging can be said to be limited by the positive electrode. Thus, the sliding of the NE profiles in Figure 2a,b does not affect the state of delithiation of the PE at the EOC, and parasitic reduction leads to slippage only of the discharge endpoint. Conversely, the large slope of the Gr profile towards the EOD singlehandedly determines when cell discharging will end, causing discharge to be limited by the NE. As a consequence, the NE will fix the discharge endpoint, regardless of the shifts experienced by the PE profiles due to parasitic oxidation in Figure 2d,e.

An important observation is that the dynamic explained above relies heavily on features of the voltage profile of Gr, which causes charging to be PE-limited and discharging to be NE-limited. The present work was motivated by the observation that, when electrode properties deviate from these conditions, the neat separation of $I_{red}$ and $I_{ox}$ using endpoint slippages no longer holds. Such properties appear in many systems of interest. While we will use Si-based NEs in most of our examples, additional cases are discussed later in this manuscript.

Consider the voltage profile of a typical Gr-free Si electrode (Figure 2g,h). No well-defined plateaus are present at high Li$^+$ content during lithiation (Figure 2g), and the profile is far less vertical than Gr towards the end of delithiation (Figure 2h), lowering the extent by which a single electrode "limits" the determination of a half-cycle. Due to the lack of a plateau, a decrease in the Li$^+$ content of the NE at the EOC due to parasitic reduction causes the "terminal" NE potential to increase. For a fixed cell voltage cutoff, the EOC must then occur at equally higher PE potentials, leading to further delithiation of the positive electrode. This mechanism causes parasitic reduction to also produce some slippage of the charge endpoint (Figure 2i). In the case of parasitic oxidation, the lower slope of the Si profile at the EOD (Figure 2k) leads to finite differences in the terminal Li$^+$ content of the NE as the cell ages, producing some discharge endpoint slippage (Figure 2l). All in all, for such systems, it is incorrect to assume that parasitic reduction and oxidation will each be quantifiable by slippage of one of the endpoints.

In the present work, we analyze these exceptions quantitatively and propose simple expressions that can extract the true extent of parasitic reactions from endpoint slippage measurements. We also discuss how these corrections may depend on the selected voltage cutoffs and on the Si content in the electrode, and how similar adjustments may be required in the context of Na-ion batteries. We also show various cases in which corrections are not needed at all, even for Si cells, and suggest how these can be identified. Extracting meaning from experimental data is not always straightforward, and having awareness of the shortcomings of analytical methods is important to achieve an accurate characterization of aging behavior.

*Experimental and Simulation*

Cell aging was simulated using voltage profiles for the PE and NE obtained from half-cells cycled at slow rates (< C/100); the exception was LiFePO$_4$ (LFP), for which we used the open circuit curve from ref. [22]. Half-cells (2032 format) were fabricated in a dry room (dew point < -45 °F), using a disc of Celgard 2500 separator and electrolytes based on a 1.2 M LiPF$_6$ solution in a 3:7 wt:wt mixture of ethylene carbonate and ethylmethyl carbonate (Tomiyama). For cells with the Si-based electrode, the electrolyte further contained 3 wt% of fluoroethylene carbonate (Solvay). Each cell contained 40 µL of electrolyte, in large excess of the total pore volume of electrodes and separator. Electrodes were fabricated at Argonne National Laboratory's Cell Analysis, Modeling and Prototyping (CAMP) Facility. Electrode composition, loading and calendered porosity are indicated below.

- NMC811: 90% active (Targray), 5% C45 carbon (Timcal), 5% PVDF (5130, Solvay); 34.5% porosity, 59 µm coating, 15.81 mg/cm$^2$
- NMC532: 90% active (Toda), 5% C45, 5% PVDF (5130, Solvay); 33.1% porosity, 42 µm coating, 11.4 mg/cm$^2$
- Graphite: 91.83% active (SLC1506T, Superior Graphite), 2% C45, 0.17% oxalic acid, 6% PVDF (KF-9300, Kureha); 37.4% porosity, 47 µm coating, 6.49 mg/cm$^2$
- Si: 70% active (150 nm nSi/C, Paraclete Energy), 15% C45, 15% LiPAA (LiOH titrate); 57% porosity, 10 µm coating, 0.85 mg/cm$^2$

Hard carbon electrodes comprised 94.95% of active material (GN-BHC-300), 0.05% single-walled carbon nanotubes (Tuball) and 5% PVDF (KF-7200, Kureha). The electrode had a loading of 10.9 mg/cm$^2$ and was calendered to 110 µm (38.4% porosity). Testing was performed in coin-cells vs. Na metal, using a glass fiber separator (Whatman) and 1 M NaPF$_6$ solution in 1:1 wt:wt ethylene carbonate and diethyl carbonate.

Voltage profiles for Si-graphite blends with various compositions were generated by combining profiles for both materials, weighted by the desired ratios of capacity contribution. It was assumed that all electrodes had 5% of binder and that the effective capacity of Si and Gr were 2500 and 350 mAh/g, respectively.

The simulation of aging was performed using Python and was made with respect to the SOC scale of the unaged positive electrode (as used in ref. [5]). We considered the cells to be at an arbitrary point after formation, where the initial state of the NE is found by scaling it according to the desired negative/positive capacity ratio (0.9 to 1.1, depending on the case) and then applying a lateral offset according to the assumed formation losses (0.15). When using the unaged PE SOC scale as reference, parasitic reduction can be represented by shifting the NE profile, whereas the opposite occurs for parasitic oxidation (see Figure 2). Voltage profiles for full-cells were obtained from the point-by-point subtraction of the NE from the PE profiles and imposing the assumed voltage cutoffs.

Simulating the effects of parasitic reactions (such as shown in Figure 2) is commonly done in an *ex post facto* manner, where the resulting cell profile embodies the cumulative effects of prior aging while assuming that no aging happens at the cycle being represented. This is justifiable, as aging occurs at minute rates in real cells and its consequences are usually assessed through sporadic slow-rate cycles. We opted to use a different approach and simulate aging *as it happens*, which facilitates the visualization of the changes in endpoints when few successive cycles with exaggerated aging rates are assumed. Importantly, this method does not introduce new trends and leads to similar EOC and EOD endpoints as the more traditional approach. Simulation of aging was achieved by distributing the desired shift throughout the entire voltage profile being modified. For instance, the shift is multiplied by zero at the beginning of a half-cycle and by increasing values at ensuing data points, until reaching the final desired offset as the voltage cutoff is met. The rate of aging is assumed to be constant for a given half-cycle. This effectively deforms the voltage profiles, but correctly captures cell behavior at both the beginning and ending of all half-cycles. Aging rates are reported throughout the manuscript for all figures, and are applied at all represented cycles, including the first one.

*Results and Discussion*

*Quantitative correlation between endpoint slippage and side-reactions.* Let us revisit Figure 2a-c, assuming that, at each half-cycle, the capacity consumed by parasitic reduction is given by $q_{red}$. During both charging and discharging, these parasitic processes cause the voltage profile of the NE to move away from that of the PE, each by $q_{red}$ (Figure 2a,b). At the EOD, the NE will have experienced a cumulative shift equal to $2q_{red}$. Since in Gr cells discharging is generally limited by the polarization of the NE, parasitic reduction will lead the discharge endpoint to slip by this same $2q_{red}$ (Figure 2c). Hence, the slippage equals the total amount of capacity consumed by parasitic reduction within each full-cycle.

A similar exercise can be done for oxidation side-reactions in Figure 2d-f. During charging, a capacity exchange of $q_{ox}$ due to parasitic oxidation will shift the voltage profile of the PE towards that of the NE (Figure 2d). Since charging is limited by the polarization of the PE, the charge endpoint of the cell will slip rightwards by $q_{ox}$. Parasitic oxidation during discharging causes a further shift of the PE profile (Figure 2e). However, since the EOD occurs when the NE polarizes, no endpoint slippage is observed; parasitic oxidation will extend both charge and discharge, but the EOD will occur at the same point in the cumulative cell capacity axis. Interestingly, since this latter oxidation occurs after the EOC, it will affect the charge endpoint *not* in the present cycle, but in the next one. Indeed, the charge endpoint at an arbitrary cycle $k$ will have slipped with respect to that in cycle $k-1$ by $q_{ox}$ during discharge $k-1$ and another $q_{ox}$ during charge $k$ (Figure 2f). In summary, when Gr is the negative electrode, it is approximately correct to consider that the charge and discharge endpoint slippages *for each cycle* can be described by

$$D_{slip} = q_{red,c} + q_{red,d} \qquad (1)$$

$$C_{slip} = q_{ox,c} + q_{ox,d,-1} \qquad (2)$$

where the indexes $c$ and $d$ indicate the appropriate half-cycle (i.e., charge and discharge, respectively), and *-1* denotes that the quantity derives from the preceding cycle. When $q_{y,c} = q_{y,d} = q_y$ (for $y = ox, red$), we have

$$D_{slip} = 2q_{red} \qquad (3)$$

$$C_{slip} = 2q_{ox} \qquad (4)$$

Recall from the discussion of Figure 2g-l that the equations above may not apply for Si cells, as the translation of parasitic capacities into endpoints slippages is distorted by the shape of the voltage profile of Si. In ref. [1] we derived expressions for how oxidation and reduction side-reactions will affect cell capacity after accounting for these shapes. In Section S1 of the Supplementary Material we use these relationships to describe endpoint slippages, showing that more general expressions that can apply to all cases in Figure 2 are

$$D_{slip} = (1-\lambda)\left(q_{red,c} + q_{red,d}\right) + \lambda(q_{ox,c} + q_{ox,d}) \tag{5}$$

$$C_{slip} = (1+\omega)(q_{ox,c} + q_{ox,d,-1}) - \omega(q_{red,c} + q_{red,d,-1}) \tag{6}$$

or, again assuming that the rate of parasitic processes remains constant across the relevant half-cycles,

$$D_{slip} = 2(1-\lambda)q_{red} + 2\lambda q_{ox} \tag{7}$$

$$C_{slip} = 2(1+\omega)q_{ox} - 2\omega q_{red} \tag{8}$$

Here, the terms $\lambda$ and $\omega$ account for the shapes of the voltage profiles, according to

$$\lambda \equiv \frac{\left.\frac{dU_{PE,d}}{dq}\right|_{EOD}}{\left(\left.\frac{dU_{PE,d}}{dq}\right|_{EOD} - \left.\frac{dU_{NE,d}}{dq}\right|_{EOD}\right)} \tag{9}$$

$$\omega \equiv \frac{\left.\frac{dU_{NE,c}}{dq}\right|_{EOC}}{\left(\left.\frac{dU_{PE,c}}{dq}\right|_{EOC} - \left.\frac{dU_{NE,c}}{dq}\right|_{EOC}\right)} \tag{10}$$

The derivatives are the slopes of the voltage profiles of PE and NE at the points where the cell meets the end of the relevant half-cycle. Since the cell voltage is the point-by-point difference between the electrical potentials experienced by the PE and NE, the denominators in eq. 9 and 10 are the slopes of the full-cell at the EOD and EOC, respectively. Taking that into consideration, it becomes clear that $\lambda$ describes how much the PE limits cell discharge, whereas $\omega$ describes how much the NE is limiting the charge. In Section S2 of the Supplementary Material we analyze the coefficients of equations (7) and (8), and show that the contributions of $q_{red}$ and $q_{ox}$ to each endpoint slippage are weighted by the extent at which the NE and the PE, respectively, are limiting the relevant half-cycle.

For the Gr cells in Figure 2, $\lambda \approx \omega \approx 0$, and equations (7) and (8) simplify to equations (3) and (4), respectively. Refer to Section S1 of the Supplementary Material for the sign convention assumed for the derivatives; also note that, within this convention, $\omega \geq 0 \geq \lambda$.

Equations (7) and (8) describe quantitatively what we have inferred from Figure 2g-l: depending on the shape of the voltage profile of the NE, $q_{red}$ and $q_{ox}$ can generate slippage of both endpoints. The equations also suggest that, for a given cycle, the slippages will depend on the total extent of side-reactions across the relevant half-cycles, regardless of how they are partitioned. That is, only $q_{red,c} + q_{red,d}$ matters, regardless of the values of $q_{red,c}$ and $q_{red,d}$. We provide direct examples of that in Figure S2 for a NMC811 vs Gr cell.

Equations (7) and (8) constitute a system of two equations containing two variables ($q_{red}$ and $q_{ox}$). All other quantities ($\omega$, $\lambda$ and the endpoint slippages per cycle) can be determined experimentally. Solving this system will result in the true parasitic capacities experienced by the cell at each cycle. In the following sections we provide examples of that process in simulated systems, illustrating the required steps and the magnitude of the distortions that can be expected for many systems of interest.

***Extracting $q_{red}$ and $q_{ox}$ from endpoint slippages.*** Figure 3 shows simulated cells that will be used in most of the examples throughout this work. The figure contains voltage profiles for the PE and NE during charge and discharge of simulated full-cells, at an arbitrary point after formation but before any simulated aging is applied. All profiles are represented along the SOC axis of the unaged PE, and the vertical lines in Figure 3a-h indicate the EOC and EOD, as appropriate. Figure 3i,j show the initial values of $\omega$ and $\lambda$, respectively. Since $\omega$ and $\lambda$ express how much a given electrode will limit a half-cycle termination, their magnitude will depend on the identity of both electrodes. Since LFP reaches the EOD at a plateau, discharging is entirely limited by the NE and $\lambda = 0$, even for Si. Similarly, the higher slope of NMC811 at lower SOCs causes the cell to present higher $\lambda$ than the NMC532 one when paired with Si.

Let us now analyze what happens when reduction side-reactions occurring at the constant rate of 0.005 per half-cycle (in PE SOC units) are imposed onto each of these cells (Figure 4a). Plots of full-cell voltage profiles in a cumulative capacity scale are shown in Figure S3. The capacity loss experienced by the Gr cell and the LFP vs. Si cell tracks well with $2q_{red}$ (black dashed line). For the two types of NMC vs. Si cells, however, the measurable capacity fade largely underestimates the occurrence of side-reactions. This deviation is caused by the *reservoir effect*, which we discuss in detail in refs. [1] and [2]. Briefly, this mismatch occurs when the range of potentials experienced by the PE and NE in successive cycles vary as a consequence of aging. Consumption of electrons by parasitic reduction implies that the PE cannot be restored to its initial Li$^+$ content when the cell is discharged, raising the terminal potential of the NMC PE. Since the cell discharges to a fixed cutoff voltage, the terminal NE potential must rise by the same extent. Due to the lower slope of the voltage profile of Si (than Gr) at the EOD, this change in potential causes a non-negligible amount of delithiation of the NE, supplying extra electrons to the PE and partially offsetting the losses to side-reactions. In the case of LFP vs. Si, the flat voltage profile of the PE at the EOD will pin the NE to a same terminal potential regardless of the rates of side-reactions, leading to good correspondence between capacity measurements and the extent of parasitic processes. In other words, differently from Gr, a non-negligible portion of the Li$^+$ stored in Si only becomes accessible at high potentials. As the NMC cell ages, progressive increases in

NE potential at the EOD will release some of these Li$^+$ back to the PE, causing the illusion of stability.

The cumulative discharge and charge endpoint slippages computed for these cases are shown in Figure 5a and Figure 5d, respectively, and trends are similar to observed for capacity: endpoint slippages can capture aging well for the Gr and the LFP cells, but poorly for the NMC vs. Si ones. If one were to trust endpoint slippages, the inferred parasitic capacities for NMC811 vs Si would underestimate $q_{red}$ by ~22% and detect a non-negligible $q_{ox}$ (equal to ~9% of $q_{red}$) where there is none. Fortunately, this can be remedied with the use of equations (7) and (8).

The first step to solve the system of equations is to collect the values of $\lambda$ and $\omega$, and of the endpoint slippages $D_{slip}$ and $C_{slip}$ *for each cycle* (Figure 5b,e). Solving the equations will then result in reliable values for $q_{red}$ and $q_{ox}$ (Figure 5c,f). Note that the corrected capacities per cycle obtained from the equations are summed and presented in their cumulative form in Figure 5c,f, mimicking how endpoint slippages are commonly presented in the literature.

We also provide an example where only parasitic oxidation is present (Figure 6). Oxidation side-reactions inject electrons into the cell and can cause capacity gains (Figure 4b) for as long as the electrolyte damage does not limit cell performance. Once more, significant deviations are seen in NMC vs Si systems, as changes in the range of potentials experienced by the electrodes constrains the accessibility to the extra electrons during cycling. Discharge endpoint slippage can be detected to a significant extent even when $q_{red} = 0$ (Figure 6d). This "phantom slippage" tends to be larger for discharge than charge, since the magnitude of $\lambda$ is generally larger than that of $\omega$ (Figure 3i,j).

While the examples above consider cases where a single parasitic processes is present and at a constant rate, the equations apply equally well to scenarios that can better represent the complexities of experimental data. Figure 7 shows a NMC811 vs Si cell in which both parasitic oxidation and reduction are present, with $q_{red}$ decaying quadratically and $q_{ox}$ growing to the 2.5 power. Figure 7a shows the changes in capacity that occur as a result of these side-reactions, along with the cumulative LLI (with $LLI = -2|q_{red} - q_{ox}|$). Due to the reservoir effect, the offset between LLI and the capacity fade is an astounding ~30.6% of the total applied LLI . The measurable endpoint slippages also misrepresent the true aging rates, with discharge underestimating and charge exaggerating the values of $q_{red}$ and $q_{ox}$, respectively (Figure 7b). Nevertheless, solving the system of equations lead to the correct determination of parasitic capacities (Figure 7c). As indicated in equation (6), the charge endpoint slippage depends on the parasitic capacities accrued during the preceding discharge half-cycle, as indicated in the legend in Figure 7c; the trace for $q_{ox,c} + q_{ox,d}$ would slightly deviate from the one presented, since the rate of aging varies from cycle to cycle.

After this initial discussion about the problems one can encounter when relying on endpoint slippages for quantifying parasitic reactions, and how they can be remedied, there are a few things worth mentioning. First, as we can see in Figure 4, Figure 5 and Figure 6, the measurability of aging is good for Gr and for LFP cells (regardless of the counter electrode, as long as the PE limits the EOC and the NE limits the EOD), and they do not require assistance from equations (7) and (8). Small deviations are seen for Gr due to the non-zero values of $\lambda$ and $\omega$ (Figure 3), but these can be neglected without major consequences to cell diagnosis.

A second point of note concerns the determination of $\lambda$ and $\omega$. In the examples above the calculations were performed using values determined at each cycle for all cells. While these parameters can be derived from experiments (e.g., using a reference electrode) and/or analyses (e.g., dV/dQ or dQ/dV fitting), that can be laborious. In most cases we have considered to date, the variation in $\lambda$ and $\omega$ with aging was sufficiently small as to be neglected without significant tradeoffs in accuracy. An example is shown in Figure S4 for a NMC811 vs Si cell experiencing parasitic reduction. Solving the system of equations using $\lambda$ and $\omega$ determined at every cycle (Figure S4a) or only using their initial values (Figure S4b) led to fairly comparable results, even with $\lambda$ varying by as much as ~0.1. Simple simulations of aged profiles (such as we do here) can be useful in estimating the range of values that can be assumed by these parameters, helping identify when such approximation is appropriate.

Finally, while the examples we discuss here consider that slippages are calculated from successive cycles (with each enduring excessive aging for ease of representation), that is not a requirement for using the equations. Figure S5 displays a case in which we simulated aging for 100 cycles and used the equations in two manners, either measuring $D_{slip}$ and $C_{slip}$ at every cycle or cumulatively every twenty cycles. Here, all slippage accrued during the middle 19 cycles are combined and used to solve the equations. The latter is equivalent to analyzing only the reference performance tests (RPTs), which are typically collected at slower rates along desired aging conditions, both in cycle and calendar life testing. Figure S5 shows that these two methods lead to the same results, both for charge and discharge endpoint slippages. The equations we present here are valid as long as $\lambda$ and $\omega$ remain constant for the cycles being considered. As we discussed above and demonstrated in Figure S4, errors can be relatively small even when some variation in the magnitude of these parameters exist. The ability to focus on RPTs and neglect other cycles can be useful in cases where it is desirable to determine $\lambda$ and $\omega$ at various points of the dataset.

*Effect of voltage cutoff in Si cells.* It is clear from the examples and equations above that the inability of endpoint slippages to inform on the rate of side-reactions depends on the shape of the voltage profiles of the PE and NE. Since Gr has a rather steep profile at low Li$^+$ content (Figure 3b), cells experiencing a depth of discharge (DOD) of ~1 will have very similar values of $\lambda$, regardless of the selected voltage cutoffs. That is not true for many Si-rich electrodes (Figure 3d). The smoother delithiation profile of Si enables cells to be planned with a variety of effective NE potential windows. In fact, typical Si electrodes may exhibit as much as ~25% of their total delithiation capacity above 500 mV vs Li/Li$^+$, and 5-10% above 700 mV vs Li/Li$^+$, and it is unlikely that they will ever be fully delithiated when cells achieve full discharge. All in all, a nominal cell DOD of 1 can reflect completely different states (and $\lambda$ values), depending on the choice of voltage cutoff.

To demonstrate that, we simulated a NMC811 vs. Si cell and analyzed its behavior when cycled between 4.15 V and three different discharge cutoff voltages: 2.7, 3.0 and 3.2 V. Figure 8a shows the voltage profile of the Si NE, with the vertical lines indicating the Li$^+$ content corresponding to each voltage limit; profiles become steeper at decreasing cutoffs. We then simulated aging for all these cases, assuming that only reduction side-reactions were taking place. Values of $\lambda$ at all cycles are shown in Figure 8b, indicating that the variability in their magnitude depend on the selected operational conditions, being significant for 3.0 V for negligible at 2.7 V. Naturally, such variation reflects changes in the shape of the voltage profiles at the EOD as the cells age. Capacity losses for all cases as a consequence of parasitic reduction are shown in Figure 8c and, as expected from the reservoir effect, always underestimated the true levels of $q_{red}$, with the gap being proportional to $\lambda$. A similar observation applies to the discharge endpoint slippages (Figure 8d), with the slippage for the cell cycled to 3.2 V missing a remarkable ~40% of the side-reactions. In contrast, deviations are small when the cell is discharged to 2.7 V, and corrections using equations (7) and (8) become less necessary. There have been reports that elevated levels of delithiation may be detrimental to some Si systems,[23-26] which may ultimately drive the decision of the discharge cutoff to be adopted. Note that the specific shape of the voltage profile of the full-cell depends on various factors, and hence the conclusions we draw for each cutoff may not be directly transferrable to other cases. In other words, testing cells to 2.7 V may result in more or in less deviations than what we show here depending on material sources, NE/PE capacity ratio, prelithiation, testing history etc. In Figure 8e,f we included the initial charge and discharge profiles, respectively, for the various cases we simulated. Such profiles provide a better reference when considering whether corrections of endpoint slippage data are needed: unless the EOD occurs with a near-vertical slope (as the 2.7 V trace in Figure 8f), deviations are likely to be expected.

*How much Si is too much?* Aging can be directly assessed through capacity and endpoint slippage measurements for Gr cells, but often not for Si. Many commercial high-energy cells have explored Si-Gr mixtures in the NE, prompting our interest in investigating at which point the deviations we discuss here would become relevant. To achieve that, we generated composite voltage profiles for systems with varying proportions of the two components, ranging from pure Gr to pure Si. Figure 9a contains delithiation voltage profiles for the various simulated cases, with the vertical line indicating the EOD at an assumed cell discharge cutoff of 3 V. The legend labels indicate the fraction of Si in the weight of the active materials in the electrode. Based on the findings of the previous section, Figure 9b shows the same series of Si-Gr composites, now in the case of the cells being discharged to 2.7 V. Comparing the two panels, it becomes clear that, a lower discharge voltages, all cases tend to resemble pure Gr at the EOD. This can also be seen in Figure 9c, which presents the values of $\lambda$ as a function of the Si content for both scenarios. At a 2.7 V discharge, values vary little as a function of Si content, and always approach that obtained for pure Gr. At a 3 V cutoff, however, $\lambda$ increases very rapidly at low Si content before plateauing at 30-40 wt%. Under these testing conditions, even small amounts of Si (e.g., 5 wt%) may be sufficient for significant distortions to be observed in the endpoint slippage measurements, requiring corrections to be made to achieve an accurate analysis.

***Other systems of interest.*** While Si electrodes have been the focal point of our examples, the thoughts we discussed here are equally relevant to several other scenarios. In this section, we explore two additional cases of relevance.

The first one is graphite. While we have established that Gr enables direct determination of $q_{red}$ and $q_{ox}$ from experimental data, that is true only when cells are fully discharged. As we have discussed before in the context of capacity measurements (ref. [2]), at most other DODs Gr will end the discharge at a plateau, causing the half-cycle to become highly limited by the PE and making $\lambda \to 1$. In Figure 10a we show the voltage profile for the delithiation of a Gr electrode, with vertical lines indicating the points at which the NE would reach the EOD for the indicated discharge cutoffs in a cell vs. NMC811. Considering that only reduction side-reactions are present, the simulated discharge endpoint slippages are shown in Figure 10b. At 3.6 V, the evolution resembles that of some Si systems we have discussed above. At 3.7 V, the EOD occurs at a plateau and the discharge endpoint remains stationary, providing no direct information about aging. In these cases, one would need to use equations (7) and (8) to accurately describe the aging of the cells. Although these conditions can be easily avoided in laboratory testing, they may occur in more realistic duty cycles.

Another system worth highlighting is hard carbon, which has been widely investigated as a NE for Na-ion batteries.[13] The voltage profile of hard carbon during desodiation generally presents three zones (Figure 10c): i) a nearly flat portion at high $Na^+$ content; ii) a slanted region at lower $Na^+$ content; iii) a nearly-vertical domain as the NE reaches complete desodiation. Inspecting this voltage profile, it becomes clear that, at the EOD, the voltage profile of hard carbon resembles that of Si, and so it can be expected that much of the distortions in capacity and endpoint slippage will also apply to these systems.

*Conclusions.*

Being able to quantify the extent of parasitic oxidation and reduction is of vital importance to evaluate new electrolyte compositions and to properly characterize cell aging. As we discussed here, tracking the endpoint slippages can provide accurate information about these side-reactions in graphite cells, but may fall short of describing Si and other systems of interest. The source of this issue is the shape of the voltage profiles of the electrodes. Endpoint slippages are able to decouple oxidation and reduction essentially because Gr has a plateau at high $Li^+$ content (making charging PE-limited) and a nearly vertical profile at low $Li^+$ content (making discharging NE-limited). At least one of these features is missing from several other NE materials, causing endpoint slippages to carry some contribution from both types of side-reactions and preventing direct inference from testing data.

By accounting for the contribution of the shape of the voltage profiles of both PE and NE to the EOC and EOD, we derived expressions enabling us to decouple these parasitic reactions in experimental data. The shapes are considered through two parameters, $\lambda$ and $\omega$, which represent how much the PE is limiting discharging and how much the NE is limiting charging, respectively. In NMC vs. Gr cells, both parameters are equal to zero, but they can assume larger values for other electrode pairs. In the examples we discussed here, using $\lambda$ and $\omega$ to correct the endpoint slippage helped avoid errors in determining the extent of parasitic reactions that could be as high as ~40%. These distortions, nevertheless, may disappear depending on the voltage range experienced by a given electrode. We discussed some of these cases, showing that, even in Si cells, it may be possible to analyze endpoint slippages without additional corrections, once certain conditions are met.

One interesting aspect of this work is that the mathematical framework we used here to generalize the application of endpoint slippages to all systems is the same we used to generalize the approach from Tornheim & O'Hanlon in ref. [1]. In other words, manipulation of a same set of equations led to the descript of the two known methods to quantifying parasitic reactions, suggesting that they are essentially equivalent. At a future work we will validate this empirically by applying both methods to study the reactivity of various electrolyte compositions.

All the analysis done here neglects contributions of losses of accessible capacity in either electrode to the endpoint slippages. This can incur in additional error, especially in systems like Si, where cyclic stresses and dilation may be present. A detailed quantification of these effects will be provided in a future manuscript.

*Acknowledgements*

This research was supported by the U.S. Department of Energy's Vehicle Technologies Office under the Silicon Consortium Project, directed by Brian Cunningham, Thomas Do, Nicolas Eidson

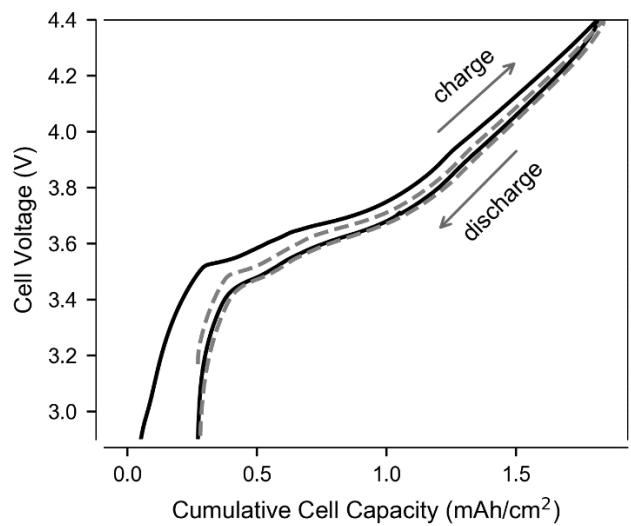

Figure 1. Experimental full-cell data represented along a cumulative capacity axis. Discharging is plotted with decreasing capacity values. Charging half-cycles are represented with onset at the point where the preceding discharging ended. The second cycle in the figure is shown in dashed gray lines.

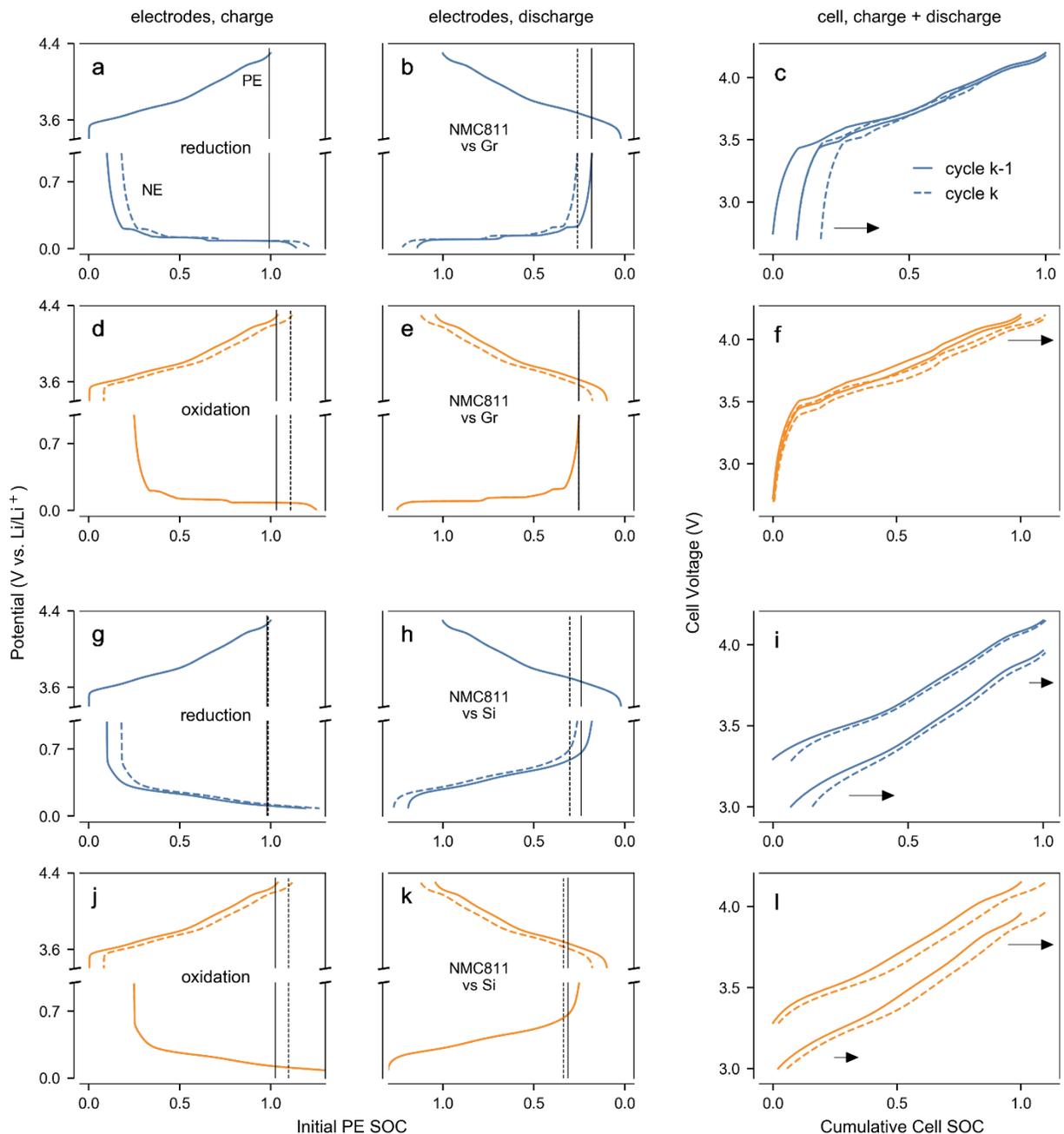

Figure 2. Voltage profiles for electrodes and cells for NMC811 vs graphite (a-f) and NMC811 vs Si (g-l). Blue and orange curves indicate cases where only parasitic reduction or oxidation are assumed, respectively. Cell profiles are obtained from the difference between the positive and the negative electrode profiles, and applying the desired voltage cutoffs. Dashed lines indicate the second cycle being represented (see legend in panel c). Vertical black lines indicate the SOC where the voltage cutoff is met for each half-cycle (with the more aged cycle represented in dashed lines). Electrode profiles are plotted along the SOC scale of the unaged positive electrode.

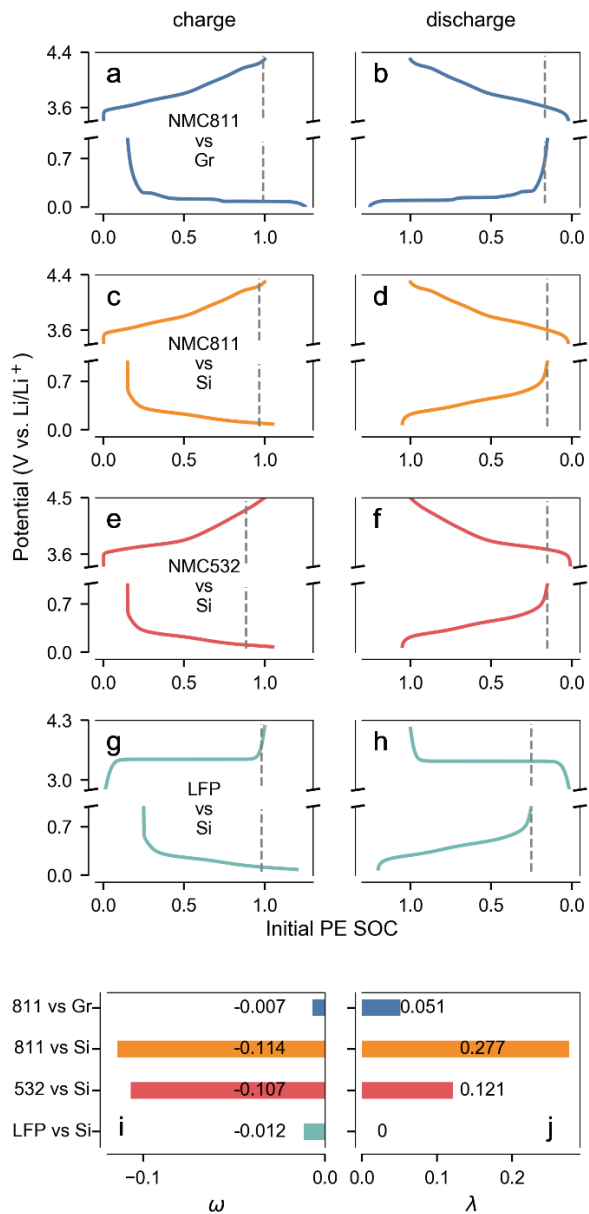

Figure 3. Voltage profiles for electrodes used to simulate cells of interest used in this work: a,b) NMC811 vs graphite (3.0-4.2 V); c,d) NMC811 vs Si (3.0-4.15 V); e,f) NMC532 vs Si (3.0-4.2 V); g,h) LFP vs Si (2.6-3.65 V). The vertical dashed lines indicate the points at which electrodes reach the end of charge (a,c,e,g) and of discharge (b,d,f,h) in simulated unaged cells. From the slopes of the voltage profiles at the EOC and EOD, the parameters $\omega$ (i) and $\lambda$ (j) are calculated.

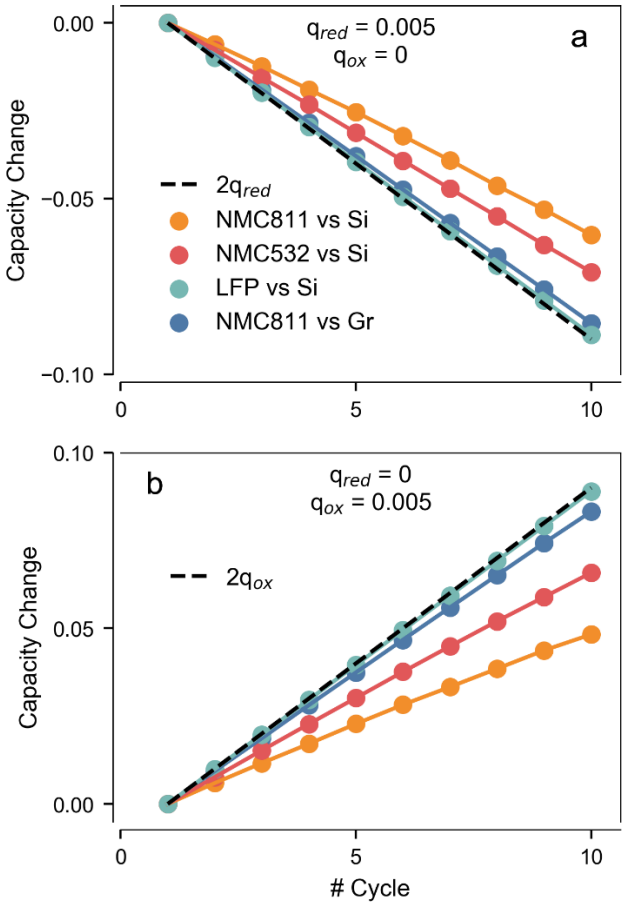

Figure 4. Capacity changes observed due to aging in simulated full-cells. A rate of 0.005 of parasitic reduction (a) or oxidation (b) per half-cycle is considered in each case. The dashed black line indicates the total capacity consumed by side-reactions, for reference. Both aging rates and capacity changes are expressed in the SOC scale of the unaged positive electrode.

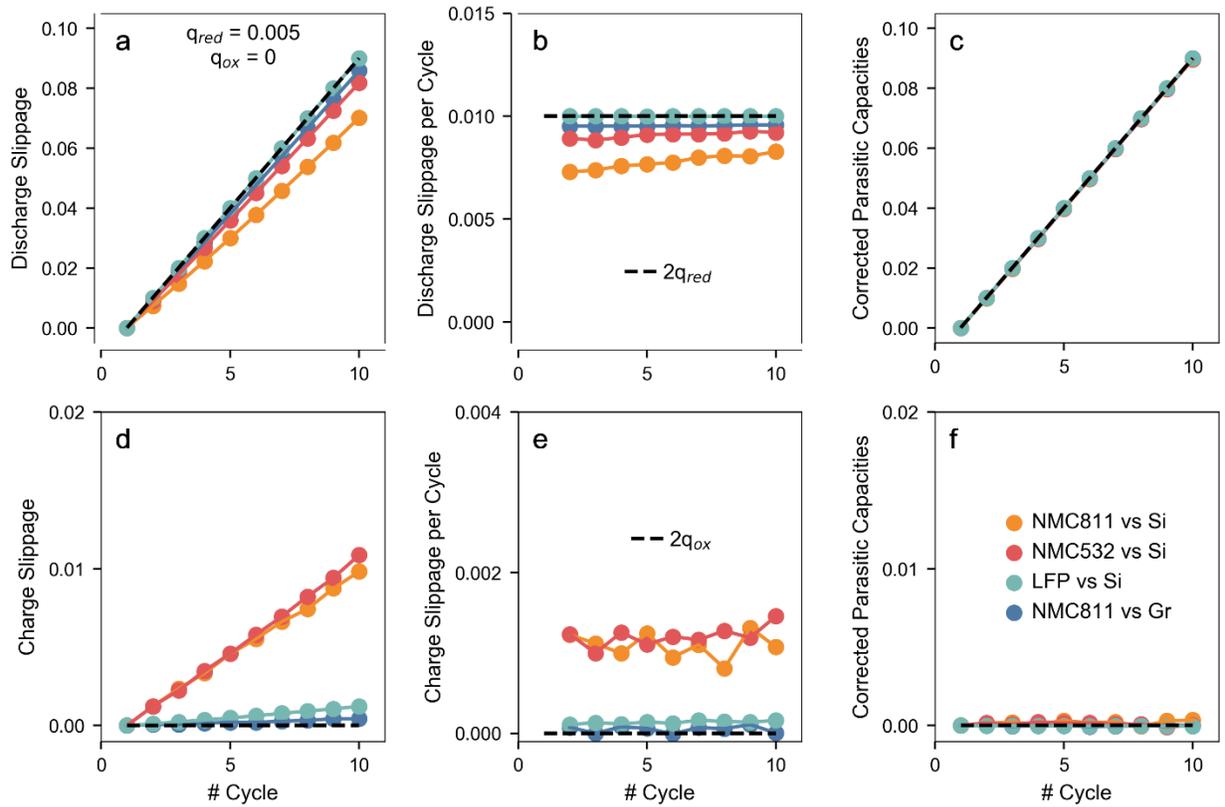

Figure 5. Simulating the effect of parasitic reduction in full-cells. a,d) Cumulative endpoint slippage; b,e) Endpoint slippage per cycle; c,f) Corrected values of parasitic capacities obtained using equations (7) and (8) (reduction in *panel c* and oxidation in *panel f*). Information for discharge is shown in a-c and for charge in d-f. The dashed black lines indicate the total capacity consumed by the appropriate side-reaction. It is assumed that only parasitic reduction is present, at a rate of 0.005 per half-cycle. Slippage, capacities and aging rates are in units of unaged PE SOC.

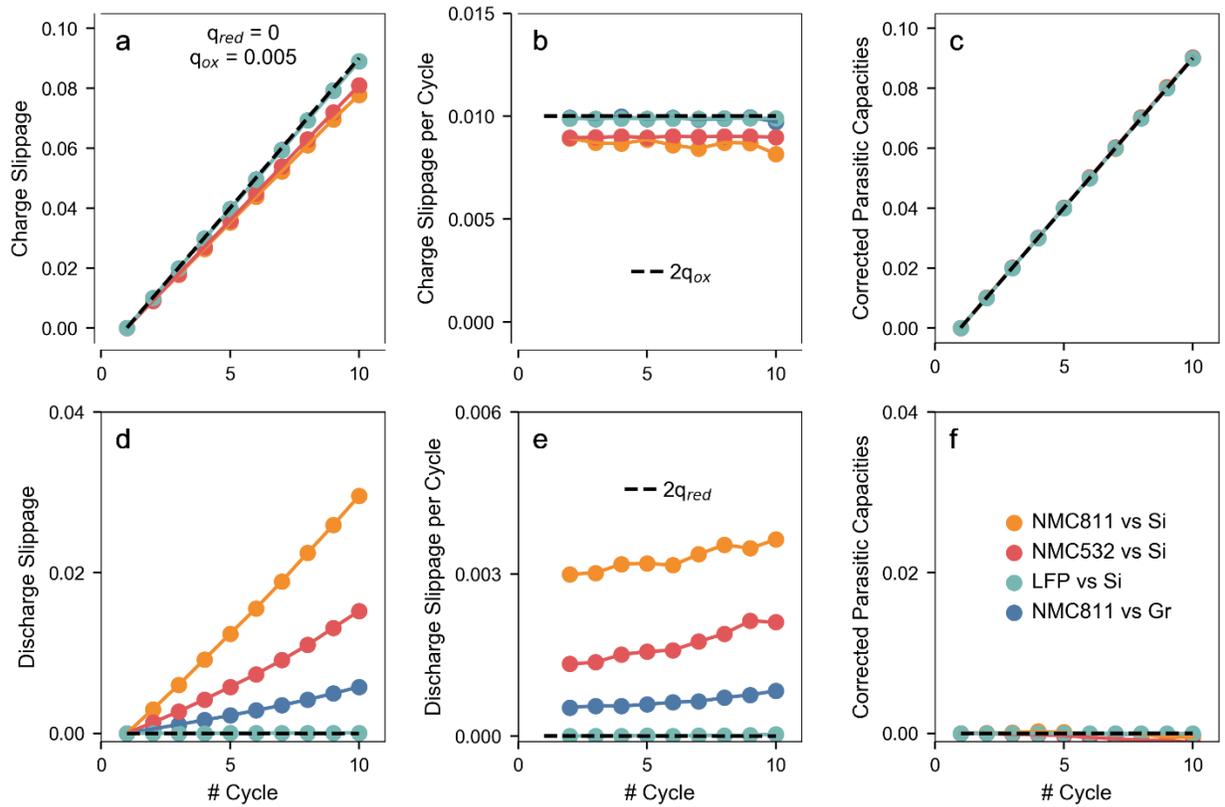

Figure 6. Simulating the effect of parasitic oxidation in full-cells. a,d) Cumulative endpoint slippage; b,e) Endpoint slippage per cycle; c,f) Corrected values of parasitic capacities obtained using equations (7) and (8) (oxidation in *panel c* and reduction in *panel f*). Information for charge is shown in a-c and for discharge in d-f. The dashed black lines indicate the total capacity consumed by the appropriate side-reaction. It is assumed that only parasitic oxidation is present, at a rate of 0.005 per half-cycle. Slippage, capacities and aging rates are in units of unaged PE SOC.

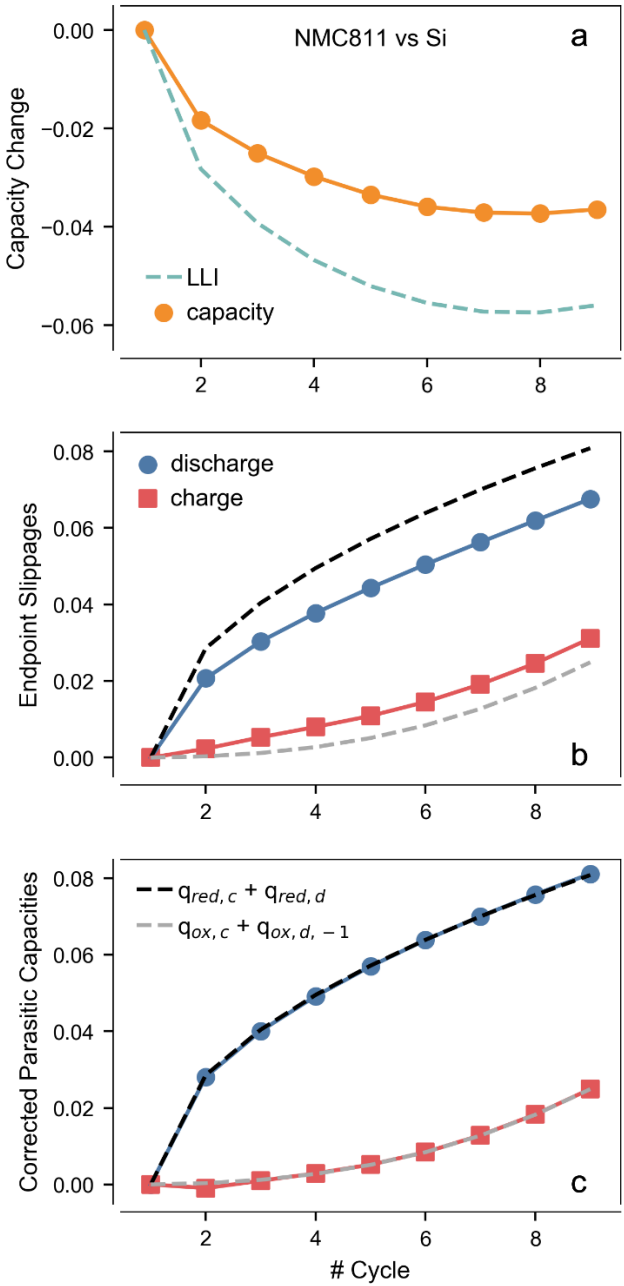

Figure 7. Side-reactions occurring at non-uniform rates in a NMC811 vs Si cell. a) Capacity change and loss of Li$^+$ inventory (LLI) as a result of simultaneous oxidation and reduction side-reactions. LLI is the difference between the cumulative $2q_{red}$ and $2q_{ox}$. The gap between the visible fade and the true electron consumption by side reactions is due to the reservoir effect. b) Endpoint slippages; and c) Corrected parasitic capacities obtained using equations (5-8). The dashed lines in b,c show the cumulative capacity consumed by the side-reactions indicated in the legend in *panel c*. The color code from *panel b* also applies to *panel c*. The equations can successfully infer the true rates of aging even in complex parasitic profiles.

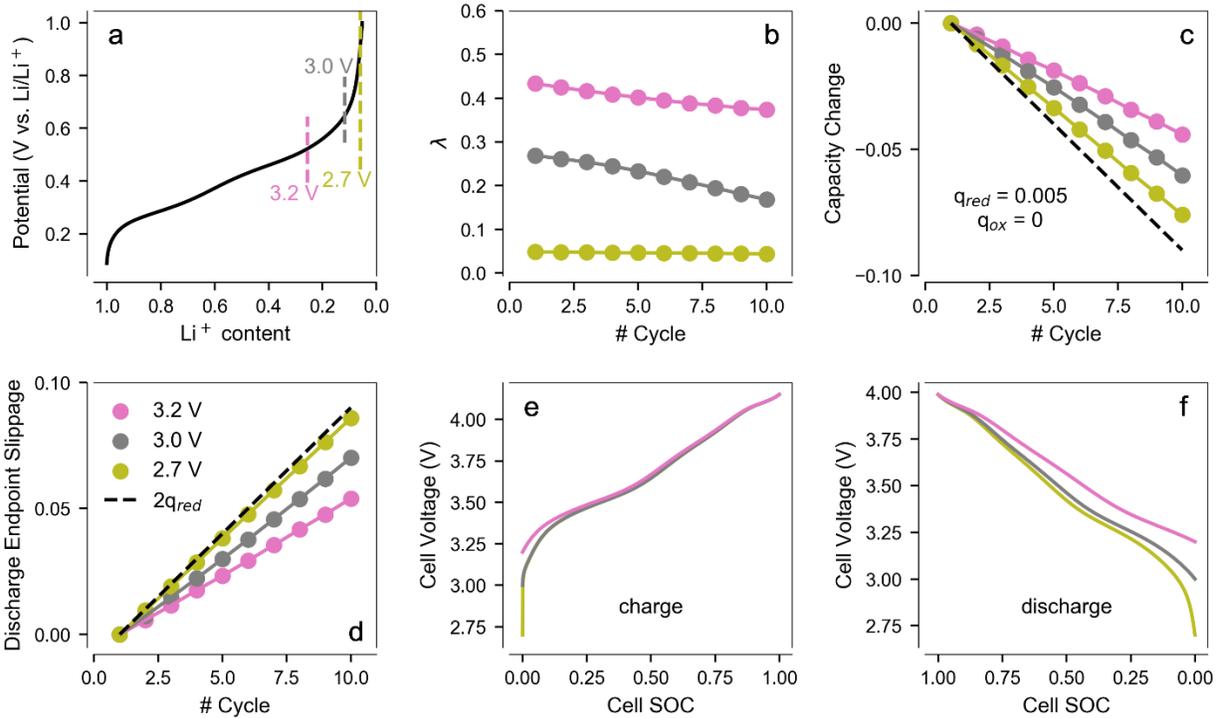

Figure 8. NMC811 vs Si cells discharged to 2.7, 3.0 and 3.2 V. a) Voltage profile during Si delithiation, with lines showing the points where the indicated discharge cutoffs would occur; b) Values of the parameter $\lambda$ as the simulated cells age; c) Capacity change; d) Discharge endpoint slippage; e) Charge profile; and f) Discharge profile of full-cells at the beginning of life. It is assumed that only parasitic reduction is present, at a rate of 0.005 per half-cycle. The dashed black line indicates the cumulative capacity consumed by these side-reactions. Slippage, capacities and aging rates are in units of unaged PE SOC.

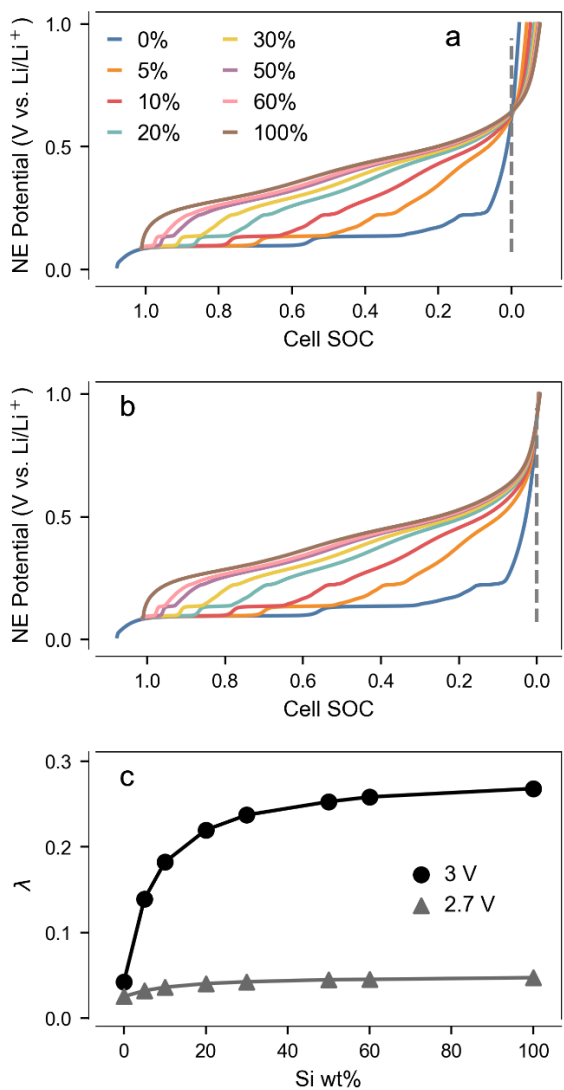

Figure 9. Simulated Si-graphite electrodes with varying Si content. a) Voltage profiles during the delithiation of Si-graphite blends when cells vs NMC811 are discharged to 3.0 V. Profiles are represented along the scale of cell SOC, and thus portions extending beyond 1 and 0 are not actively utilized in the cycle. b) Like *panel a*, but for a discharge cutoff of 2.7 V. The vertical dashed lines indicate the end of discharge. c) Initial values of $\lambda$ as a function of Si content for both assumed cutoffs. The effects of the shape of the voltage profile of Si on the measurability of aging depends strongly on the extent of delithiation. The color code in *panel a* indicates the Si content in the simulated profiles.

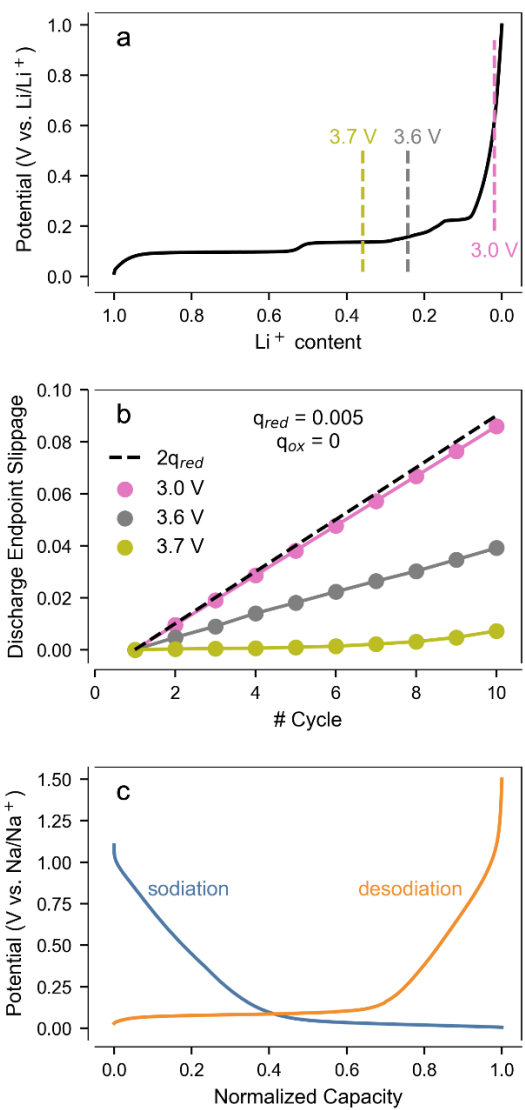

Figure 10. Additional examples of cases where direct measurements of endpoint slippages may provide incorrect information. a) Voltage profile during delithiation of a graphite electrode. Vertical lines denote the $Li^+$ content of the NE at the indicated discharge cutoffs in a cell vs NMC811. b) Discharge endpoint slippage obtained after applying 0.005 of parasitic reduction per half-cycle to cells with the discharge cutoffs shown in *panel a*. c) Voltage profiles for a hard carbon electrode in a Na-ion half-cell. The voltage profile at latter parts of desodiation have low slopes until ~1 V vs $Na/Na^+$, indicating that the measurability of aging through discharge endpoint measurements will depend strongly on the voltage cutoff of full-cells.

# Supplementary Material

## Addressing limitations of the endpoint slippage analysis

Marco-Tulio F. Rodrigues

Chemical Sciences and Engineering Division, Argonne National Laboratory, Lemont, IL, USA

**Contact:** marco@anl.gov

## Section S1. Derivation

Definitions:

- PE = positive electrode
- NE = negative electrode
- EOC = end of charge
- EOD = end of discharge
- SOC = state of charge
- $Q_c$ and $Q_d$ are the measured full-cell charge and discharge capacities, respectively
- $Q_{BOC}$ is the amount of capacity that would be exchanged during the charge half-cycle in the absence of parasitic processes within that half-cycle (i.e., the ideal accessible Li$^+$ inventory at the beginning of charge)
- $Q_{BOD}$ is the amount of capacity that would be exchanged during the discharge half-cycle in the absence of parasitic processes within that half-cycle (i.e., the ideal accessible Li$^+$ inventory at BOD).
- $I_j$ is the constant current used at a given half-cycle (j = c,d for charge and discharge, respectively).
- $\tau_j$ is the duration of the half-cycle (j = c,d for charge and discharge, respectively).
- $I_{ox,j}$ is the time-averaged rate of oxidation side reactions at the half-cycle j (j = c,d for charge and discharge, respectively); $I_{ox,j} = \tau_j^{-1} \int_0^{\tau_j} \gamma_{ox,j}(t)dt$, where $\gamma_{ox,j}(t)$ is the function describing the changes in rate of oxidation side reactions as a function of time (and, implicitly, of electrode SOC/potential). $I_{ox,j}$ is always > 0.

- $I_{red,j}$ is the time-averaged rate of reduction side reactions at the half-cycle j (j = c,d for charge and discharge, respectively); $I_{red,j} = \tau_j^{-1} \int_0^{\tau_j} \gamma_{red,j}(t)dt$, where $\gamma_{red,j}(t)$ is the function describing the changes in rate of reduction side reactions as a function of time (and, implicitly, of electrode SOC/potential). $I_{red,j}$ is always > 0.

- $\left.\frac{dU_{i,j}}{dq}\right|_k$ is the slope of the voltage profile of the electrode i (i = PE, NE for positive and negative electrodes, respectively) during the half-cycle j (j = c,d for charge and discharge, respectively) around the point k (k = EOC and EOD). It is assumed that the slope remains constant within the small variation in electrode SOC as the cells loses/gains capacity within a single cycle. (That is, the slope of the anode profile is assumed to be the same at the end of two successive discharges despite the occurrence of a small shift in the electrode potential at the EOD.)

- $\left.\frac{dU_{i,j}}{dq}\right|_k$ is measured within the frame of reference in which other calculations are performed. For example, most of the analyses in the present manuscript are made with respect to the PE SOC scale, and hence electrodes need to be represented in that scale prior to computing the derivative. Similarly, when analyzing experimental data, voltage profiles for the electrodes must first be represented at the relevant scale.

- $\left.\frac{dU_{i,c}}{dq}\right|_k = -\left.\frac{dU_{i,d}}{dq}\right|_k$, as we are comparing the electrode profiles during charge and discharge.

- $\left.\frac{dU_{PE,c}}{dq}\right|_{EOC}$ and $\left.\frac{dU_{NE,d}}{dq}\right|_{EOD}$ are > 0; $\left.\frac{dU_{PE,d}}{dq}\right|_{EOD}$ and $\left.\frac{dU_{NE,c}}{dq}\right|_{EOC}$ are < 0.

- Electrode potentials at the end of a half-cycle are referred to as "terminal potentials".

In the Supplementary Material of ref. S1, we have derived expressions for the measurable charge and discharge capacities as a function of the rates of parasitic reactions $I_{red}$ and $I_{ox}$, and of the shapes of the voltage profiles (as embodied by the parameters $\lambda$ and $\omega$ defined in the main manuscript). We refer the reader to this derivation for additional context, as we will use some outcomes of that previous work to determine expressions for the endpoint slippages.

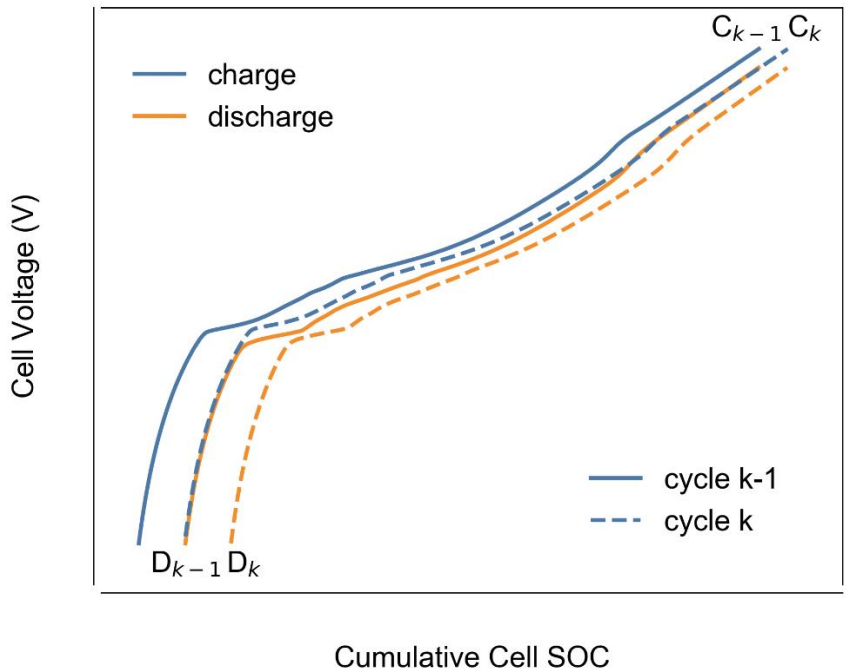

*Figure S1.* *Simulated voltage profiles for two successive cycles ($k-1$ and $k$) for a hypothetical cell. C and D denote the charge and discharge endpoints at the indicated cycles. The discontinuity between successive half-cycles is caused by the use of slow-rate half-cell data in the simulations and is equivalent to having a rest in between half-cycles. This feature does not interfere with the representation of aging.*

Figure S1 shows voltage profiles for a cell during successive charge and discharge half-cycles at an arbitrary cycle $k$. The onset of charge coincides with the endpoint of the preceding discharge ($D_{k-1}$). Thus, the discharge endpoint slippage within the cycle of consideration $D_{slip,k}$ is simply the charge capacity minus the discharge capacity:

$$D_{slip,k} = D_k - D_{k-1} = Q_{c,k} - Q_{d,k} \qquad (S1)$$

Similarly, it can be seen from Figure S1 that the charge endpoint slippage between cycles $k-1$ and $k$, $C_{slip,k}$, is given by

$$C_{slip,k} = C_k - C_{k-1} = Q_{c,k} - Q_{d,k-1} \qquad (S2)$$

Knowing $Q_c$, $Q_d$ and $Q_{d,-1}$ we can determine $D_{slip}$ and $C_{slip}$ (dropping the index $k$ henceforward to simplify the notation).

From ref. S1, we have that

$$Q_c = Q_{BOC} + \tau_c[I_{ox,c}(1+\omega) - \omega I_{red,c}] \qquad (S3)$$

$$Q_d = Q_{BOD} - \tau_d[I_{red,d}(1-\lambda) + \lambda I_{ox,d}] \qquad (S4)$$

$$Q_{d,-1} = Q_{BOC} - \tau_{d,-1}[I_{ox,d,-1}(1+\omega) - \omega I_{red,d,-1}] \qquad (S5)$$

$$Q_{BOD} = Q_c - \tau_c[I_{red,c}(1-\lambda) + \lambda I_{ox,c}] \qquad (S6)$$

Using these definitions to express $D_{slip}$, we get

$$D_{slip} = Q_c - Q_d$$

$$= Q_{BOC} + \tau_c[I_{ox,c}(1+\omega) - \omega I_{red,c}]$$

$$- \{Q_{BOD} - \tau_d[I_{red,d}(1-\lambda) + \lambda I_{ox,d}]\} \quad (S7)$$

$$= Q_{BOC} - Q_{BOD} + \tau_c[I_{ox,c}(1+\omega) - \omega I_{red,c}]$$

$$+ \tau_d[I_{red,d}(1-\lambda) + \lambda I_{ox,d}]$$

Substituting eq. S3 in eq. S6:

$$Q_{BOD} = Q_{BOC} + \tau_c[I_{ox,c}(1+\omega) - \omega I_{red,c}] - \tau_c[I_{red,c}(1-\lambda) + \lambda I_{ox,c}] \quad (S8)$$

And now substituting eq. S8 in eq. S7:

$$D_{slip} = Q_{BOC} - \{Q_{BOC} + \tau_c[I_{ox,c}(1+\omega) - \omega I_{red,c}] - \tau_c[I_{red,c}(1-\lambda) + \lambda I_{ox,c}]\}$$

$$+ \tau_c[I_{ox,c}(1+\omega) - \omega I_{red,c}] + \tau_d[I_{red,d}(1-\lambda) + \lambda I_{ox,d}]$$

$$= -\tau_c[I_{ox,c}(1+\omega) - \omega I_{red,c}] + \tau_c[I_{red,c}(1-\lambda) + \lambda I_{ox,c}] \quad (S9)$$

$$+ \tau_c[I_{ox,c}(1+\omega) - \omega I_{red,c}] + \tau_d[I_{red,d}(1-\lambda) + \lambda I_{ox,d}] =$$

$$= \tau_c[I_{red,c}(1-\lambda) + \lambda I_{ox,c}] + \tau_d[I_{red,d}(1-\lambda) + \lambda I_{ox,d}]$$

Considering that the capacity $q_{b,j} = \tau_j I_{b,j}$ for $j = c, d$ and $b = red, ox$:

$$D_{slip} = q_{red,c}(1-\lambda) + \lambda q_{ox,c} + q_{red,d}(1-\lambda) + \lambda q_{ox,d} \quad (S10)$$

Finally, combining $q_{red}$ and $q_{ox}$ terms:

$$D_{slip} = (1 - \lambda)(q_{red,c} + q_{red,d}) + \lambda(q_{ox,c} + q_{ox,d}) \tag{S11}$$

The expression above is eq. (5) of the main manuscript. We can follow a similar process to determine $C_{slip}$:

$$\begin{aligned} C_{slip} &= Q_c - Q_{d,-1} \\ &= Q_{BOC} + \tau_c[I_{ox,c}(1 + \omega) - \omega I_{red,c}] \\ &\quad - \{Q_{BOC} - \tau_{d,-1}[I_{ox,d,-1}(1 + \omega) - \omega I_{red,d,-1}]\} \\ &= \tau_c[I_{ox,c}(1 + \omega) - \omega I_{red,c}] + \tau_{d,-1}[I_{ox,d,-1}(1 + \omega) - \omega I_{red,d,-1}] \\ &= q_{ox,c}(1 + \omega) - \omega q_{red,c} + q_{ox,d,-1}(1 + \omega) - \omega q_{red,d,-1} \end{aligned} \tag{S12}$$

Combining $q_{red}$ and $q_{ox}$ terms:

$$C_{slip} = (1 + \omega)(q_{ox,c} + q_{ox,d,-1}) - \omega(q_{red,c} + q_{red,d,-1}) \tag{S13}$$

which is equation (6) of the main manuscript.

## Section S2. Electrode limitation fractions

The full-cell voltage is the point-by-point difference between the electrical potentials $U_n$ experienced by the PE and NE. For voltage profiles represented along the relevant capacity axis $q$:

$$U_{PE} - U_{NE} = U_{cell} \tag{S14}$$

Taking the derivative with respect to capacity, we get

$$\frac{dU_{PE}}{dq} - \frac{dU_{NE}}{dq} = \frac{dU_{cell}}{dq} \tag{S15}$$

We can then write

$$\frac{\frac{dU_{PE}}{dq}}{\frac{dU_{PE}}{dq} - \frac{dU_{NE}}{dq}} - \frac{\frac{dU_{NE}}{dq}}{\frac{dU_{PE}}{dq} - \frac{dU_{NE}}{dq}} = 1 \tag{S16}$$

Comparing the expression above with the definitions of $\lambda$ and $\omega$ in equations (9) and (10) of the main manuscript, it becomes clear that

$$\frac{\frac{dU_{PE,c}}{dq}\bigg|_{EOC}}{\frac{dU_{PE,c}}{dq}\bigg|_{EOC} - \frac{dU_{NE,c}}{dq}\bigg|_{EOC}} - \omega = 1 \rightarrow 1 + \omega = \frac{\frac{dU_{PE,c}}{dq}\bigg|_{EOC}}{\frac{dU_{PE,c}}{dq}\bigg|_{EOC} - \frac{dU_{NE,c}}{dq}\bigg|_{EOC}} \tag{S17}$$

$$\lambda - \frac{\left.\frac{dU_{NE,d}}{dq}\right|_{EOD}}{\left.\frac{dU_{PE,d}}{dq}\right|_{EOD} - \left.\frac{dU_{NE,d}}{dq}\right|_{EOD}} = 1 \rightarrow 1 - \lambda = -\frac{\left.\frac{dU_{NE,d}}{dq}\right|_{EOD}}{\left.\frac{dU_{PE,d}}{dq}\right|_{EOD} - \left.\frac{dU_{NE,d}}{dq}\right|_{EOD}} \quad (S18)$$

In our sign convention, the slope of the voltage profile of the cell and of the NE are always opposite each other (e.g., $< 0$ for the cell but $> 0$ for the NE during discharge). Hence, in equations (7) and (8) of the main manuscript, $1 - \lambda$ and $-\omega$ are $> 0$ and fundamentally express by how much the NE is limiting the EOD and EOC, respectively. The same applies to the PE and the quantities $\lambda$ and $1 + \omega$.

*Section S3. Other supporting figures*

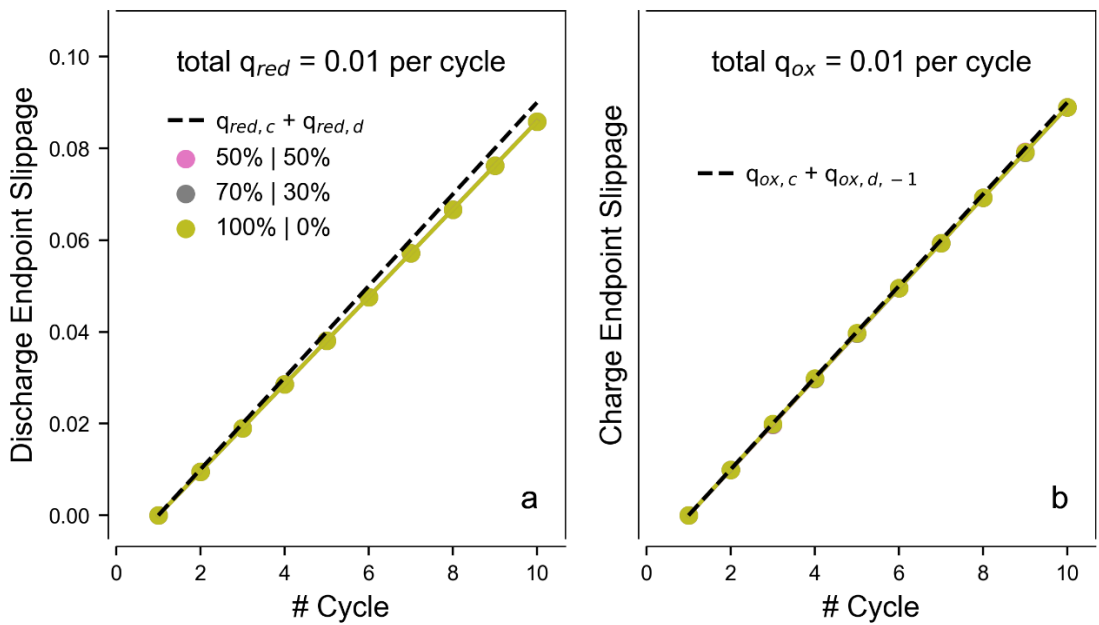

***Figure S2.*** *Analyzing the effects of the partition of side-reaction rates across half-cycles of a NMC811 vs graphite cell. Three different cases are considered for: a) 0.01 of total parasitic reduction per cycle; b) 0.01 of total oxidation per cycle. The percentages of the total parasitic capacity exchanged at each half-cycle (e.g., charge and discharge for reduction) are shown in the color code in panel a. The dashed lines indicate the cumulative parasitic capacities. Endpoint slippage depends only on the total exchanged parasitic capacity at the relevant half-cycles, regardless of how they are partitioned. All quantities are in units of unaged PE SOC.*

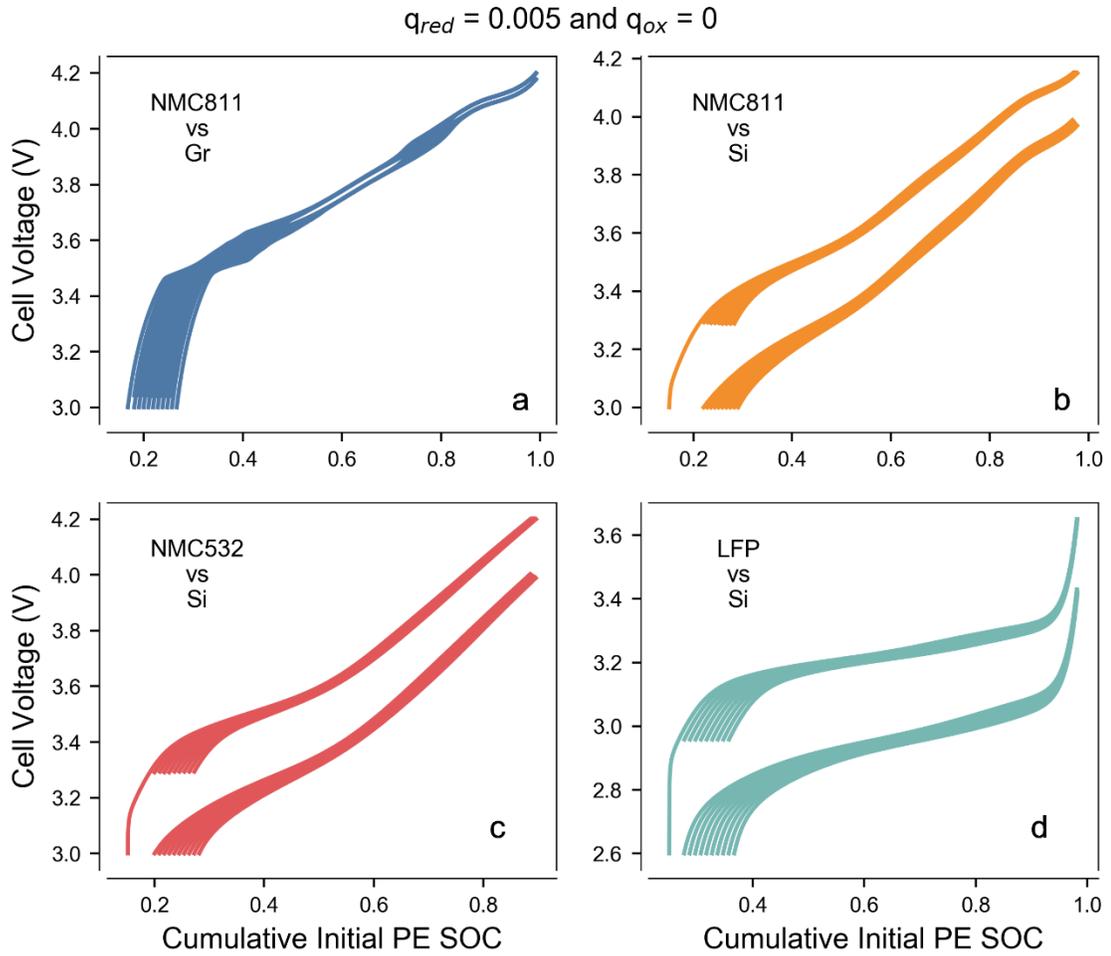

*Figure S3.* *Voltage profiles for all cells discussed in Figure 5, plotted along a cumulative capacity axis. Parasitic reduction produces visible slippage of charge endpoints in NMC vs Si cells, but not in the others.*

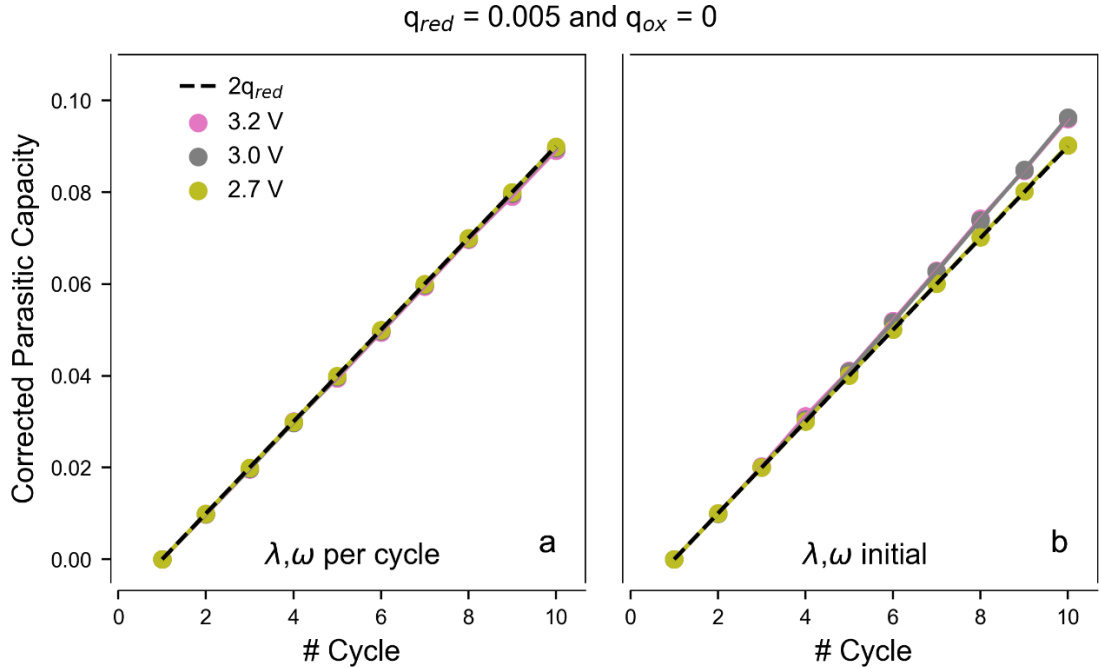

***Figure S4.*** *Corrected parasitic capacities for NMC811 vs Si cells simulated with various cutoffs and assuming a constant reduction rate of 0.005 per cycle. Cells are the same as discussed in Figure 8. Equations (7) and (8) were solved using: a) $\lambda$ and $\omega$ determined at every cycle; b) initial $\lambda$ and $\omega$ values. In many cases, the variance in the values of these parameters is small enough and allows a single set of values to be used, which can facilitate data analysis. The dashed lines indicate the cumulative parasitic capacities, and all quantities are in units of unaged PE SOC.*

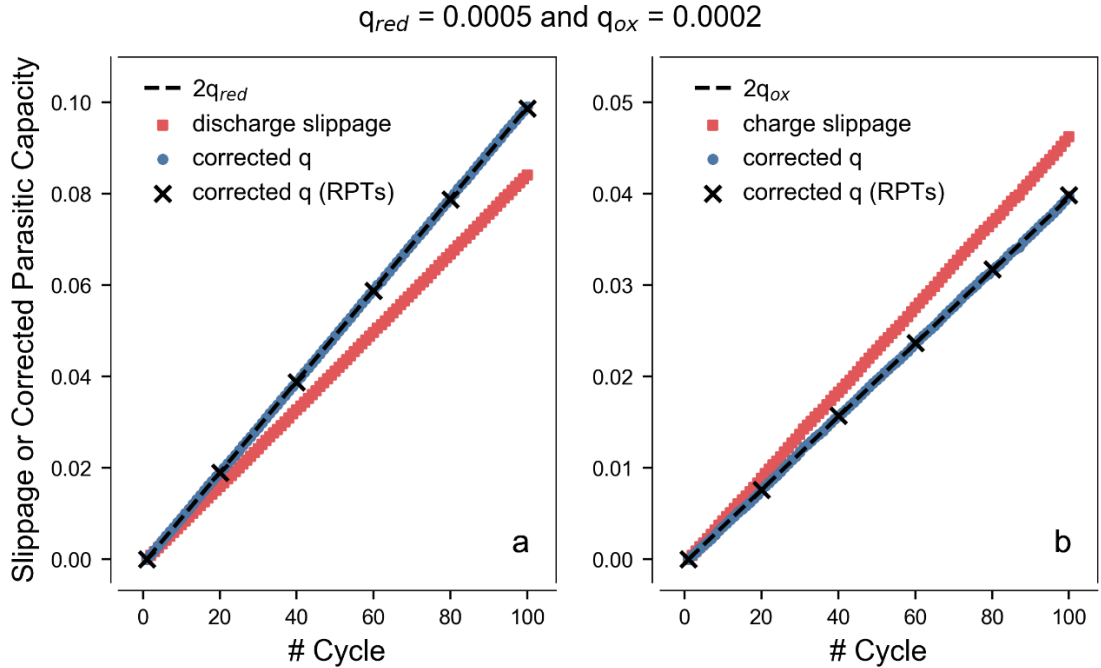

***Figure S5.*** *Using equations (7) and (8) to correct slippages at every cycle (blue) or every 20 cycles (x symbols) in a NMC811 vs Si cell. When skipping cycles, the total slippage accrued in that interval is summed and treated as a single cycle in the equations. A total of 100 cycles were simulated assuming the rate of parasitic reduction to be 0.0005 and that of oxidation to be 0.0002. Endpoint slippages are shown in red for: a) Discharge; b) Charge. The corrected parasitic capacities are nearly identical, with minute variations observed due to the use of different values for λ and ω. The dashed lines indicate the cumulative parasitic capacities, and all quantities are in units of unaged PE SOC.*

*Supporting References*